 \documentclass[onecollarge,natbib]{svjour2}
 \bibpunct{[}{]}{;}{n}{}{,} 
 \smartqed  
 \usepackage{graphicx}
 \journalname{Few-Body Systems}

\usepackage{amssymb}
\usepackage[intlimits]{amsmath}
\usepackage{amsfonts}
\usepackage{dsfont}
\usepackage{slashed}
\usepackage{units}
\usepackage{wrapfig}

      \newcommand{\conjg}[1]{\ensuremath{\hspace{1pt}\overline{\hspace{-1pt}#1\hspace{-1pt}}}\hspace{1pt}}

\usepackage[usenames]{color}
\usepackage{colortbl}

\definecolor{webgreen}{rgb}{0,0.75,0}
\definecolor{webred}{rgb}{0.75,0,0}
\definecolor{webblue}{rgb}{0,0,0.75}
\definecolor{darkblue}{rgb}{0,0,0.6}
\definecolor{darkgreen}{rgb}{0,0.5,0.5}
\definecolor{darkpurple}{rgb}{0.5,0,0.5}
\definecolor{darkorange}{rgb}{1,0.5,0}
\definecolor{darkgrey}{rgb}{0.4,0.4,0.4}
\definecolor{lgray}{rgb}{0.95,0.95,0.95}
\definecolor{lgreen}{rgb}{0.95,1.00,0.90}
\definecolor{lred}{rgb}{1.00,0.90,0.80}
\definecolor{lblue}{rgb}{0.2,0.35,1.00}
\definecolor{shadecolor}{rgb}{1.00,0.92,0.82}

\usepackage[colorlinks=true,linkcolor=lblue,citecolor=lblue,urlcolor=lblue]{hyperref}

\begin{document}

 \title{Progress in the calculation of nucleon transition form factors \thanks{This work is supported by the German Science Foundation DFG under project number DFG TR-16.}
 }


 \author{Gernot Eichmann}


 \institute{G. Eichmann \at
               Institut f\"{u}r Theoretische Physik, Justus-Liebig-Universit\"at Giessen \\
               Heinrich-Buff-Ring 16, 35392 Giessen, Germany \\
               Tel.: +49 641 9933342\\
               Fax: +49 641 9933309\\
               \email{gernot.eichmann@theo.physik.uni-giessen.de}   }

 \date{Received: date / Accepted: date}

 \maketitle

 \begin{abstract}

       We give a brief account of the Dyson-Schwinger and Faddeev-equation approach
       and its application to nucleon resonances and their transition form factors.
       We compare the three-body with the quark-diquark approach
       and present a quark-diquark calculation for the low-lying nucleon resonances including scalar, axialvector, pseudoscalar and vector diquarks.
       We also discuss the timelike structure of transition form factors and highlight the advantages of form factors over helicity amplitudes.


 \end{abstract}

 \section{Introduction} \label{intro}

       Understanding the spectrum and structure of nucleon resonances has been among the main challenges in hadron physics ever since the discovery of the Roper resonance.
       Much experimental information on the photo- and electroexcitations of baryonic resonances has been collected at Jefferson Lab, MAMI, ELSA, and other facilities~\cite{Klempt:2009pi,Aznauryan:2011qj};
       especially Jefferson Lab with the CLAS detector in Hall B has contributed a large amount to the electroproduction world data in recent years.
       Transition form factors over a wide kinematic domain are now available for a number of nucleon resonances
       and provide a unique window into the nonperturbative structure of Quantum Chromodynamics (QCD).

       There is still an abundance of open questions concerning the nature of resonances.
       Are the three quarks in a baryon spatially equally distributed or do they cluster into diquarks?
       What is the importance of molecular components that are generated by meson-baryon interactions, and how does the `pion cloud' reveal itself in form factors?
       How are confinement and spontaneous chiral symmetry breaking manifest, and what is the microscopic origin of vector-meson dominance?

       To address these questions, it is desirable to establish a consistent microscopic description
       of $\pi$, $\rho$, nucleon and nucleon resonance properties within QCD.
       We will focus on the Dyson-Schwinger equation (DSE) framework, whose basic promise is to calculate
       such observables from the nonperturbative structure of QCD's Green functions~\cite{Roberts:1994dr}.
       While there has been progress with regard to meson spectrum and structure properties as well as nucleon and $\Delta$ form factors (among other areas),
       its application to excited baryons is still at an early stage.
       In the following we give a brief account of the approach on its way towards calculating
       nucleon resonances and transition form factors.

 \newpage

            \begin{figure*}[t!]
            \centerline{%
            \includegraphics[width=0.98\textwidth]{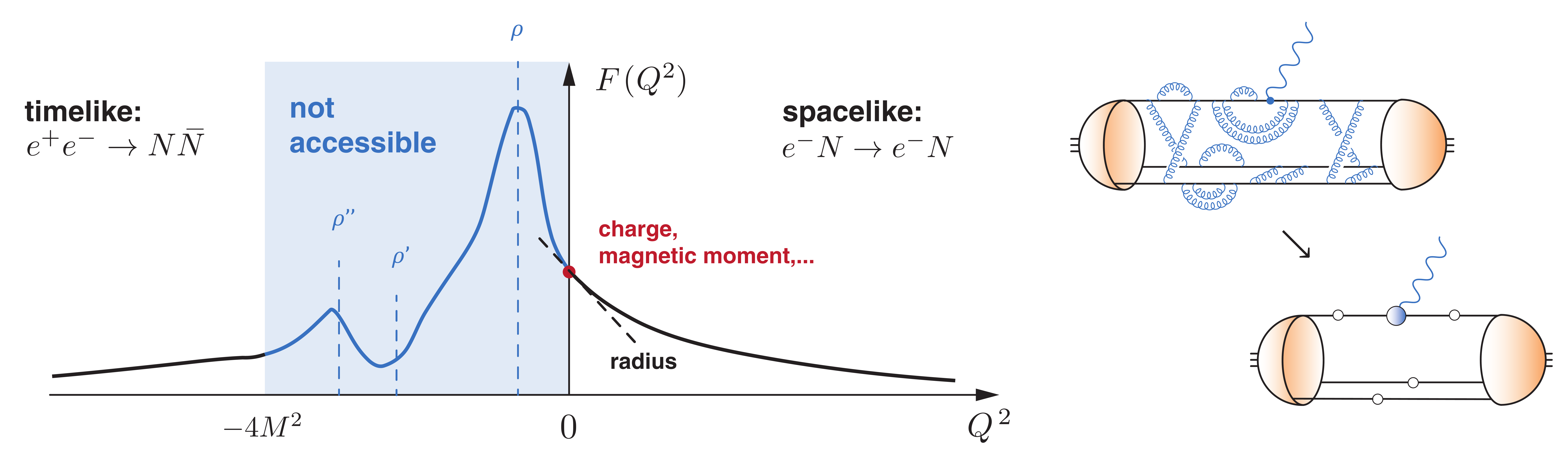}}
            \caption{\textit{Left:} Sketch of a generic form factor in the spacelike and timelike region. \textit{Right:} Representative (rainbow-ladder-like) contribution to a $NN^\star$ transition matrix element.}
            \label{fig:timelike-ffs-1}
            \end{figure*}

            \begin{figure*}[t]
            \centerline{%
            \includegraphics[width=0.75\textwidth]{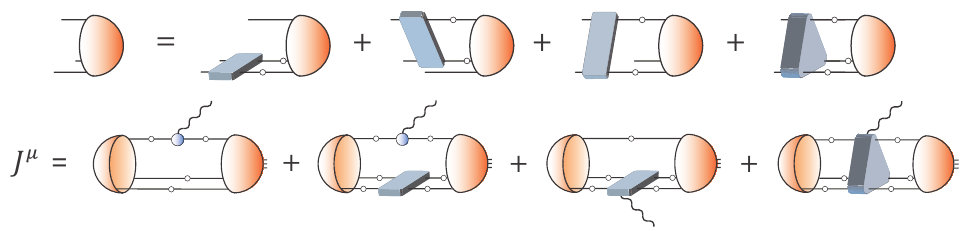}}
            \caption{Three-quark Faddeev equation (top) and electromagnetic current matrix element (bottom).}
            \label{fig:faddeev}
            \end{figure*}

 \section{The covariant Faddeev approach} \label{sec:faddeev}

             To illustrate the basic ideas, consider an electromagnetic nucleon or nucleon-to-resonance
             transition form factor. Its typical shape will be kindred to that in Fig.~\ref{fig:timelike-ffs-1}:
             its spacelike behavior ($Q^2 > 0$) is experimentally extracted from $eN$ scattering or, in the case of a resonance, pion photo- and electroproduction.
             Experimental information on timelike nucleon form factors above threshold comes from the reaction $e^+ e^-\to N\conjg{N}$ or its inverse.
             For nucleon resonances, the region below threshold is indirectly accessible
             via the Dalitz decays $N^\ast\to Ne^+e^-$ that contribute to the dilepton `cocktail' in $NN$ and heavy-ion collisions.
             The characteristic features in this regime are the vector-meson bumps that are produced when the photon fluctuates into $\rho$ and $\omega$.
             The bump landscape in Fig.~\ref{fig:timelike-ffs-1} is drawn from the pion form factor, experimentally measured via $e^+ e^-\to\pi^+\pi^-$,
             where this property is exposed due to the much smaller threshold ($2m_\pi < m_\rho$).

             Is it reasonable to expect a similar behavior also for nucleon transition form factors,
             and how can one understand these features from the quark level?
             The current matrix element that encodes the form factors
             will be made of diagrams such as that in Fig.~\ref{fig:timelike-ffs-1}:
             the incoming baryon splits into its valence quarks which emit and reabsorb gluons, obtain a boost from the photon,
             and finally recombine into the outgoing baryon.
             The blobs represent the baryons' covariant Faddeev amplitudes, the quantum field-theoretical analogues of a baryon wave function.
             Fortunately, these complicated interactions can be absorbed into a few compact building blocks: the Faddeev amplitudes, the dressed quark propagator, and the dressed quark-photon vertex.
             The complete expression for the electromagnetic current matrix element is shown in Fig.~\ref{fig:faddeev};
             it also couples the photon to the two- and three-quark kernels and thereby satisfies electromagnetic gauge invariance.
             In turn, the Faddeev amplitudes must be solved from their corresponding Faddeev equations,
             which at the same time determine the masses of the baryons~\cite{Eichmann:2009qa}.

\begin{figure*}[t]
            \centerline{%
            \includegraphics[width=0.9\textwidth]{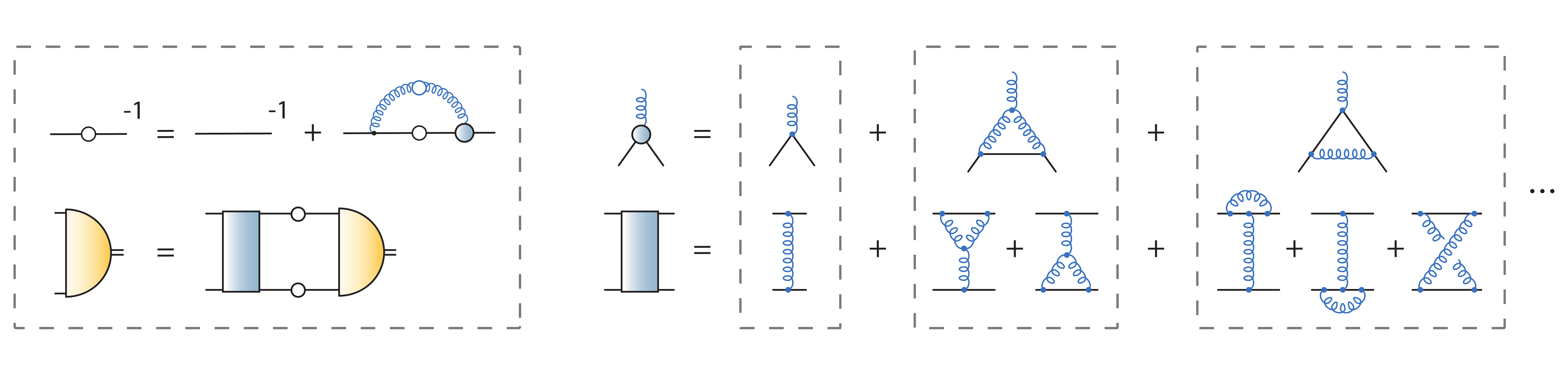}}
            \caption{Quark DSE, meson BSE, and symmetry-preserving relation between quark-gluon vertex and $q\conjg{q}$ kernel.}
            \label{fig:dse-bse}
            \end{figure*}

             The quark propagator and kernels that enter the equations are not arbitrary but related to each other.
             The quark propagator
             is determined from its Dyson-Schwinger equation (DSE) in Fig.~\ref{fig:dse-bse}.
             The resulting quark mass function becomes momentum-dependent;
             it describes the transition from the input current-quark mass at large momenta to a nonperturbative, dressed `constituent quark' mass of a few hundred MeV
             in the infrared.
             The analogous $q\bar{q}$ kernel that appears in a meson's Bethe-Salpeter equation (BSE)
             is connected to the quark propagator via chiral symmetry and the corresponding axial Ward-Takahashi identity (WTI), which pictorially
             amounts to `cutting' dressed quark lines in the DSE; see~\cite{Sanchis-Alepuz:2015tha} and references therein.
             The leading kernel contribution is a gluon exchange with a bare quark-gluon vertex  (`rainbow-ladder');
             the BSE will then produce a ladder of gluons upon iteration.

\begin{figure*}[t]
            \centerline{%
            \includegraphics[width=0.95\textwidth]{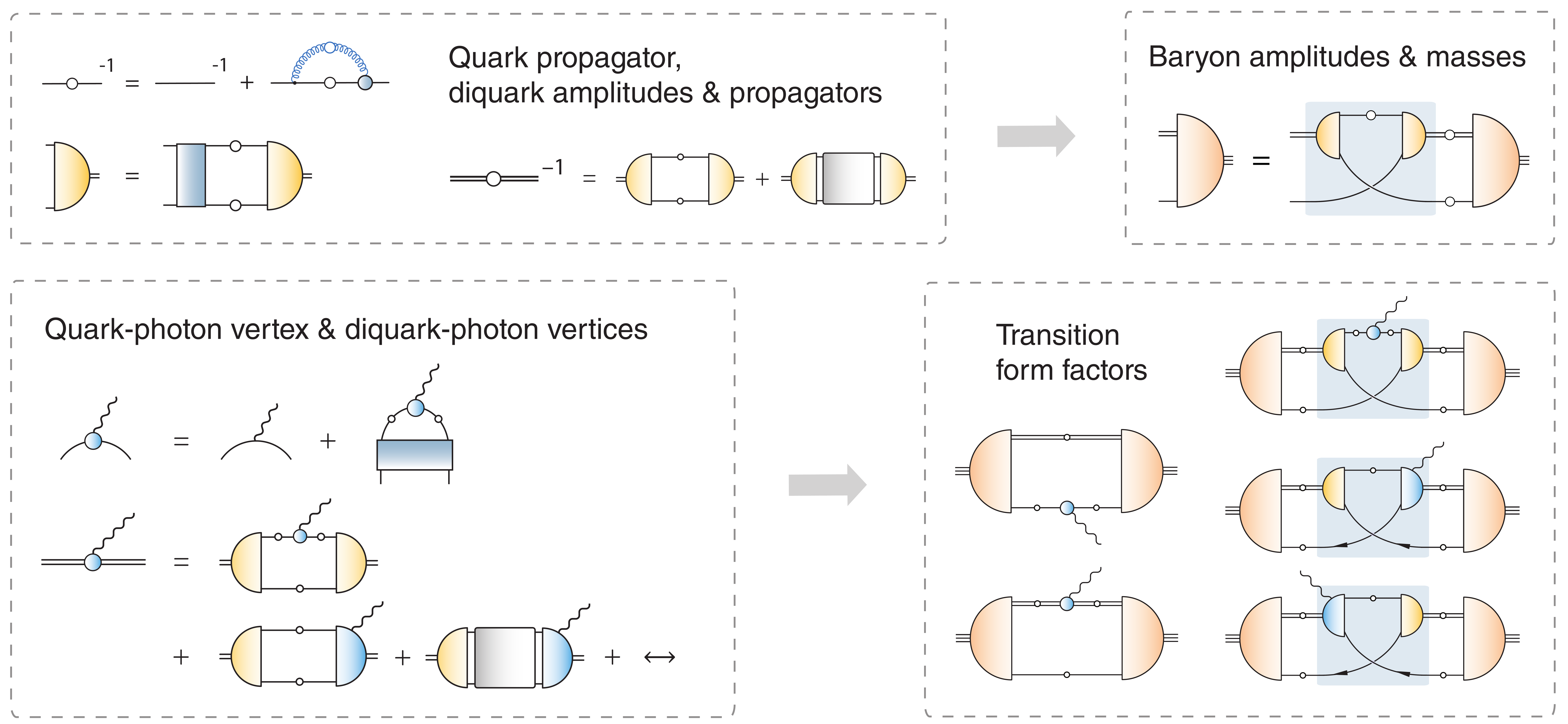}}
            \caption{Recipe for calculating transition form factors in the quark-diquark approach.
                     The quark propagator, diquark amplitudes and diquark propagators form the input of the quark-diquark BSE.
                     The current matrix element needs in addition the quark-photon vertex, the diquark-photon vertices and the seagull amplitudes.
                     The ingredients are calculated from their DSEs and BSEs in rainbow-ladder truncation.}
            \label{fig:quark-diquark}
            \end{figure*}

             In combination with spontaneous chiral symmetry breaking, the axial WTI ensures that the pion is the massless Goldstone boson in the chiral limit~\cite{Maris:1997hd}.
             Hence, if we solve the BSE for the pion from Fig.~\ref{fig:dse-bse} we will get these features for free:
             the pion mass follows the Gell-Mann-Oakes-Renner relation for small quark masses ($m_\pi^2 \sim m_q$) and vanishes in the chiral limit where the current-quark mass $m_q$ is zero.
             Another key ingredient is the quark-photon vertex: its inhomogeneous BSE (cf. Fig.~\ref{fig:quark-diquark}) features the same $q\bar{q}$ kernel,
             so the resulting vertex respects the vector WTI which is a necessary prerequisite for electromagnetic gauge invariance and thus current conservation.
             In addition, the BSE automatically generates vector-meson poles in the transverse part of the vertex~\cite{Maris:1999bh}.
             Since any form factor diagram ultimately couples the photon to the quarks through a dressed quark-photon vertex,
             this is the microscopic origin of the timelike resonance structure sketched in Fig.~\ref{fig:timelike-ffs-1}. (In rainbow-ladder
             $\rho$ and $\omega$ are bound states without a width, hence one obtains a series of poles instead of bumps.)

            \begin{wrapfigure}{R}{0.35\textwidth}
            \centerline{%
            \includegraphics[scale=0.10]{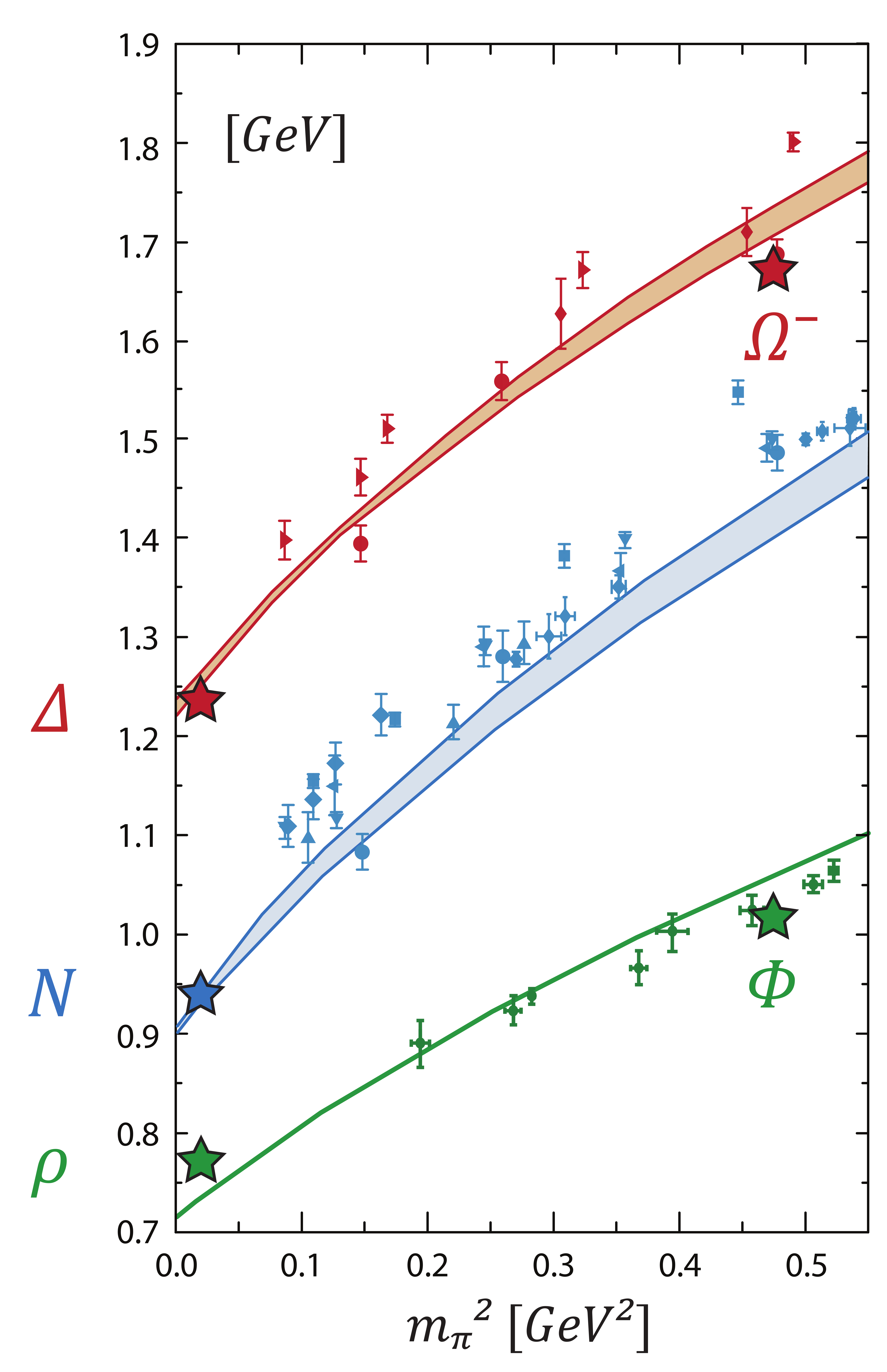}}
            \caption{$\rho-$meson~\cite{Maris:1999nt}, nucleon and $\Delta$ masses~\cite{Eichmann:2009qa}
                     calculated from their Bethe-Salpeter and Faddeev equations.
                     Stars are PDG values and symbols with error bars are lattice data (see~\cite{Eichmann:2009qa} for references).}
            \label{fig:masses}
            \end{wrapfigure}

             Apparently there is a deep underlying connection between the ingredients of these equations.
             Fig.~\ref{fig:dse-bse} also makes clear that one cannot insert a crossed-ladder term without a consistent quark-gluon vertex;
             one cannot add a confinement potential or drop spin-orbit terms by hand; etc.
             Although in principle one is equipped with an exact set of equations,
             in practice one has to make concessions due to the complexity of the problem:
             at any stage one could employ model ans\"atze instead of self-consistent solutions,
             but approximations or truncations can be chosen so that all symmetries are maintained throughout.

             Whereas the applicability of rainbow-ladder in the light-meson sector is mainly limited to pseudoscalar and vector mesons,
             baryons fare much better: the approach reproduces the octet and decuplet ground state masses within $5-10\%$~\cite{Sanchis-Alepuz:2014sca}.
             Fig.~\ref{fig:masses} shows results for the $\rho-$meson, nucleon and $\Delta$ masses as functions of $m_\pi^2$ (which is also calculated)
             compared to lattice data and experiment. The only input is the quark-gluon interaction whose model dependence is given by the bands.
             In particular, once the model scale is set to reproduce the pion decay constant, there are no further parameters or approximations and all subsequent results are predictions.

             Apart from mass spectra, a range of form factors have been calculated as well within this setup. Among them are
             nucleon, $\Delta$ and hyperon electromagnetic form factors, the $N\to\Delta\gamma$ transition, and nucleon axial form factors; see~\cite{Alkofer:2014bya} and references therein.
             All these cases exhibit good overall agreement with experimental data (where available) and also lattice results at larger pion masses, with discrepancies at low $Q^2$ where pion-cloud effects
             become important.

             We emphasize that the three-body Faddeev approach does not depend on explicit diquark degrees of freedom.
             Nevertheless, the assumption of dominant quark-quark correlations inside baryons is quite natural, and it can
             contribute much to our understanding and interpretation of nucleon transition form factors.
             In the following we will therefore review the properties of diquarks and discuss the diquark composition of nucleon resonances.

 \newpage

 \section{Nucleon resonances in the quark-diquark approach} \label{sec:ff}

             Diquarks are not observable because they carry color. Nevertheless,
             there are several observations that support a quark-diquark interpretation of baryons.
             First, the presumably leading irreducible three-body force (the three-gluon vertex that connects three quarks)
             vanishes simply due the color traces~\cite{Sanchis-Alepuz:2015tha}, which suggests that two-quark correlations are more important.
             Second, the interaction between two quarks is attractive in the color-$\mathbf{\conjg 3}$ channel.
             Neglecting three-body terms, the Faddeev equation in Fig.~\ref{fig:faddeev} can be rewritten in a more familiar form
             that depends on the $qq$ scattering matrix. Assuming that the latter is separable and can be approximated by a sum over diquark correlations,
             the Faddeev equation simplifies to a two-body problem --- the quark-diquark BSE in Fig.~\ref{fig:quark-diquark}.
             The baryon is then bound by quark exchange;
             gluons no longer appear explicitly but they are rather absorbed into the building blocks: the quark propagator, diquark amplitudes and diquark propagators.
             The electromagnetic current is constructed accordingly~\cite{Oettel:1999gc}.

             Scalar and axialvector diquarks are the lightest ones, hence they are most important for describing the positive-parity nucleon and $\Delta$ baryons.
             Diquark masses have been calculated in rainbow-ladder~\cite{Maris:2002yu} which produces actual diquark poles in the $qq$ scattering matrix.
             Although their appearance is presumably a truncation artifact~\cite{Bender:1996bb},
             it still suggests the presence of strong diquark correlations and allows one to compute diquark properties in close analogy to those of mesons from their BSEs.\footnote{Similarly,
             the four-body amplitude for a tetraquark is dominated by the lowest-lying two-body poles (the pseudoscalar mesons),
             which leads to a `meson-molecule' interpretation for the light scalar mesons~\cite{Eichmann:2015cra}.}

             Quark-diquark models yield results for a range of baryon properties that are indeed similar to those obtained from the three-body equation.
             There are essentially three variants of the approach that have been pursued in recent studies.
             One is the NJL/contact-interaction model from where qualitative statements and overall properties of the baryon spectrum have been extracted~\cite{Roberts:2011cf}.
             Another is the QCD-based model developed in~\cite{Oettel:1998bk} which implements more realistic models for the propagators, amplitudes and vertices
             and thereby also allows one to make quantitative predictions for form factors, such as in a recent calculation of the nucleon to Roper transition~\cite{Segovia:2015hra}.

\begin{figure*}[t!]
            \centerline{%
            \includegraphics[width=0.95\textwidth]{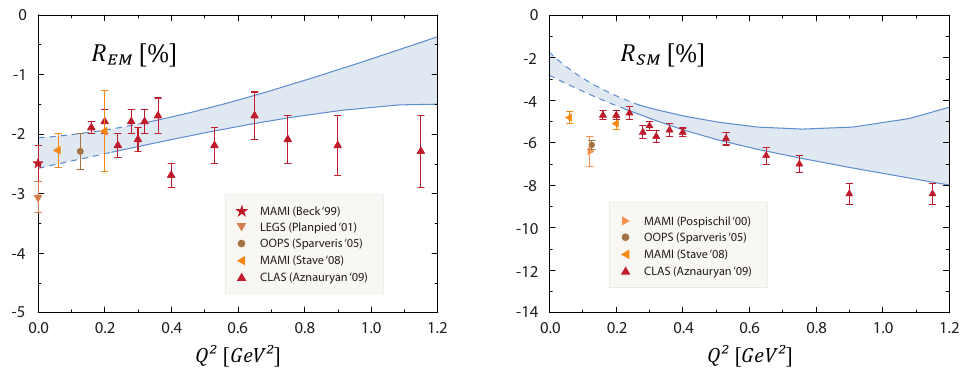}}
            \caption{$Q^2$-dependence of the electric and Coulomb quadrupole form-factor ratios $R_{EM}$ and $R_{SM}$ compared to experimental data~\cite{Eichmann:2011aa}.}
            \label{fig:ndg}
            \end{figure*}

             Here we will revisit and extend a third variant of the model, namely the approach of Ref.~\cite{Eichmann:2009zx} which is outlined in Fig.~\ref{fig:quark-diquark}.
             In that case the model input is replaced by DSE and BSE solutions, which
             provide a direct link to the microscopic building blocks since
             any modification in the fundamental quark-gluon interaction induces consistent changes up to the observables of interest.
             The calculated nucleon and $\Delta$ masses in that setup are indeed comparable to those in Fig.~\ref{fig:masses}.
             As an example we quote the $N\gamma\to\Delta$ calculation of Ref.~\cite{Eichmann:2011aa} in Fig.~\ref{fig:ndg}.
             The calculated form factor ratios $R_{EM}$ and $R_{SM}$ agree rather well with the experimental data,
             which can be partially traced to the presence of relativistic $p$ waves in the nucleon and $\Delta$ amplitudes.
             On the other hand, the magnetic dipole form factor agrees with experiment only above $Q^2 \gtrsim 1$ GeV$^2$ and deviates by $\sim 30\%$ at $Q^2=0$, presumably due to missing pion-cloud effects.
            So far the approach has been applied to nucleon and $\Delta$ ground state properties, also including their elastic form factors and the $N\to\Delta\pi$ decay~\cite{Eichmann:2009zx,Nicmorus:2010sd}.
            What are then the necessary steps to calculate negative-parity resonances such as the $N(1535)$ and $N(1520)$?
            Let us first collect some properties of the diquarks.

 \vspace{-3mm}

 \paragraph{Diquark properties.}
            Diquarks are subject to the Pauli principle, so they must be totally antisymmetric under quark exchange.
            Concerning isospin, the diquark flavor wave functions are either antisymmetric ($I=0$) or symmetric ($I=1$).
            Since color is antisymmetric, the corresponding Dirac parts must be either antisymmetric ($I=0$) or symmetric ($I=1$) as well.
            Denoting the total onshell momentum by $P$ (with $P^2=-m^2$, where $m$ is the respective diquark mass) and the relative momentum by $q$, this entails
                  \begin{equation}
                     \Gamma(q,P) = \left\{\begin{array}{rcr} -\Gamma^T(-q,P) & \dots & I=0 \\[1mm]  \Gamma^T(-q,P) & \dots & I=1 \end{array}\right.
                  \end{equation}
            for the Dirac parts of the $J^P=0^\pm$ diquark amplitudes, where $T$ is a matrix transpose.
            The same relations hold for the $J^P=1^\pm$ amplitudes with the replacement $\Gamma(q,P) \to \Gamma^\mu(q,P)$.
            This leads to the classification in Table~\ref{tab:diquarks}, where we display the two leading (s-wave) tensor structures for each case.
            The $0^\pm$ amplitudes depend on four tensors and the $1^\pm$ amplitudes on eight, but the remaining tensor components are suppressed because they
            carry relative momentum and thus higher orbital angular momentum.
            The two isospin states must differ by $\omega=q\cdot P$ to ensure the correct symmetry (the dressing functions
            are all even in $\omega$) and the appearance of $\omega$ induces further suppression.

            One can draw a close analogy between diquarks and mesons:
            each diquark amplitude can be mapped onto a respective meson by removing the charge-conjugation matrix $C$.
            The upper and lower entries in Table~\ref{tab:diquarks} then correspond to opposite C-parities for mesons.
            After taking traces, the rainbow-ladder BSEs for diquarks only differ by a factor 2 from their meson parity partners ---
            diquarks are `less bound' than mesons.
            The scalar diquarks are the lightest ones ($\sim 800$ MeV), followed by axialvector ($\sim 1$~GeV), pseudoscalar and vector diquarks.
            The meson analogues of $I(J^P)=1 (0^+)$, $0 (1^+)$ and $1 (0^-)$ are exotic and therefore
            the respective diquark masses are larger.
            In contrast to their meson counterparts, diquarks are rather sensitive to the quark-gluon interaction: in Table~\ref{tab:diquarks} we quote
            the central values for the model employed herein but the masses can vary by $100 \dots 150$ MeV in both directions~\cite{Maris:2002yu}.
            There is another consequence of the diquark-meson analogy: whereas pseudoscalar and vector mesons are well described in rainbow-ladder,
            scalar and axialvector mesons come out too light~\cite{Maris:2002yu,Krassnigg:2009zh} ---
            which can be remedied with more sophisticated truncations~\cite{Chang:2009zb}. Hence, the pseudoscalar and vector diquarks
            should inherit this behavior and produce negative-parity baryons that are also too light and acquire repulsive shifts beyond rainbow-ladder~\cite{Roberts:2011cf}.

    \renewcommand{\arraystretch}{1.4}

             \begin{table*}[t]

             \begin{equation*}
             \begin{array}{@{\quad} l @{\quad} | @{\quad} c @{\quad} | @{\quad} c  @{\quad} | @{\quad} c  @{\quad} | @{\quad} c  @{\!\;}     }

                   I \;(J^P)     & 0^+   & 1^+  &  0^-   &  1^-     \\[1mm] \hline\hline \rule{-0.0mm}{0.5cm}

                       I=0 & \gamma_5 C                                          &  \color{lblue}\omega\,\gamma^\mu C                      & C                                     & \gamma^\mu \gamma_5 C  \\
                           & i\slashed{P} \,\gamma_5 C                           &  \color{lblue}\omega\, [\gamma^\mu, i\slashed{P} ] \, C & \color{lblue}\omega\,i\slashed{P}\,C  & \color{lblue}\omega\, [\gamma^\mu, i\slashed{P} ] \, \gamma_5 C  \\[2mm]
                       \text{Mass [GeV]}       &  0.80                           &  \color{lblue}-                                         & 1.01                                  &  1.12            \\
                       \text{Meson partner} & \pi\,(0^{-+})                      &  \color{lblue}\text{exotic}\,(1^{-+})                   & \text{scalar}\,(0^{++})               &  a_1\,(1^{++})   \\[1mm]\hline \rule{-0.0mm}{0.5cm}
                       I=1 & \color{lblue} \omega\, \gamma_5 C                   &  \gamma^\mu C                                           & \color{lblue}\omega\,C                & \color{lblue}\omega\, \gamma^\mu \gamma_5 C  \\
                           & \color{lblue} \omega\,i\slashed{P} \,\gamma_5 C     &  [\gamma^\mu, i\slashed{P} ] \, C                       & i\slashed{P}\,C                       & [\gamma^\mu, i\slashed{P} ] \, \gamma_5 C \\[2mm]
                       \text{Mass [GeV]}       &  \color{lblue}-              &  1.00                                                      & \color{lblue}-                        &  1.12            \\
                       \text{Meson partner} & \color{lblue}\text{exotic}\,(0^{--}) &  \rho\,(1^{--})                                       & \color{lblue} (0^{+-})                &  b_1\,(1^{+-})   \\[1mm] \hline

             \end{array}
             \vspace{1mm}
             \end{equation*}

               \caption{$I(J^P)$ quantum numbers for the lowest-lying diquarks and their leading tensor structures.
                $C=\gamma_4 \gamma_2$ is the charge-conjugation matrix.
                The entries in light (blue) color are suppressed due to factors $\omega=q\cdot P$.
                The rainbow-ladder masses are quoted together with the respective meson parity partners.}
               \label{tab:diquarks}

             \end{table*}

            \begin{figure*}[t!]
            \centerline{%
            \includegraphics[width=0.74\textwidth]{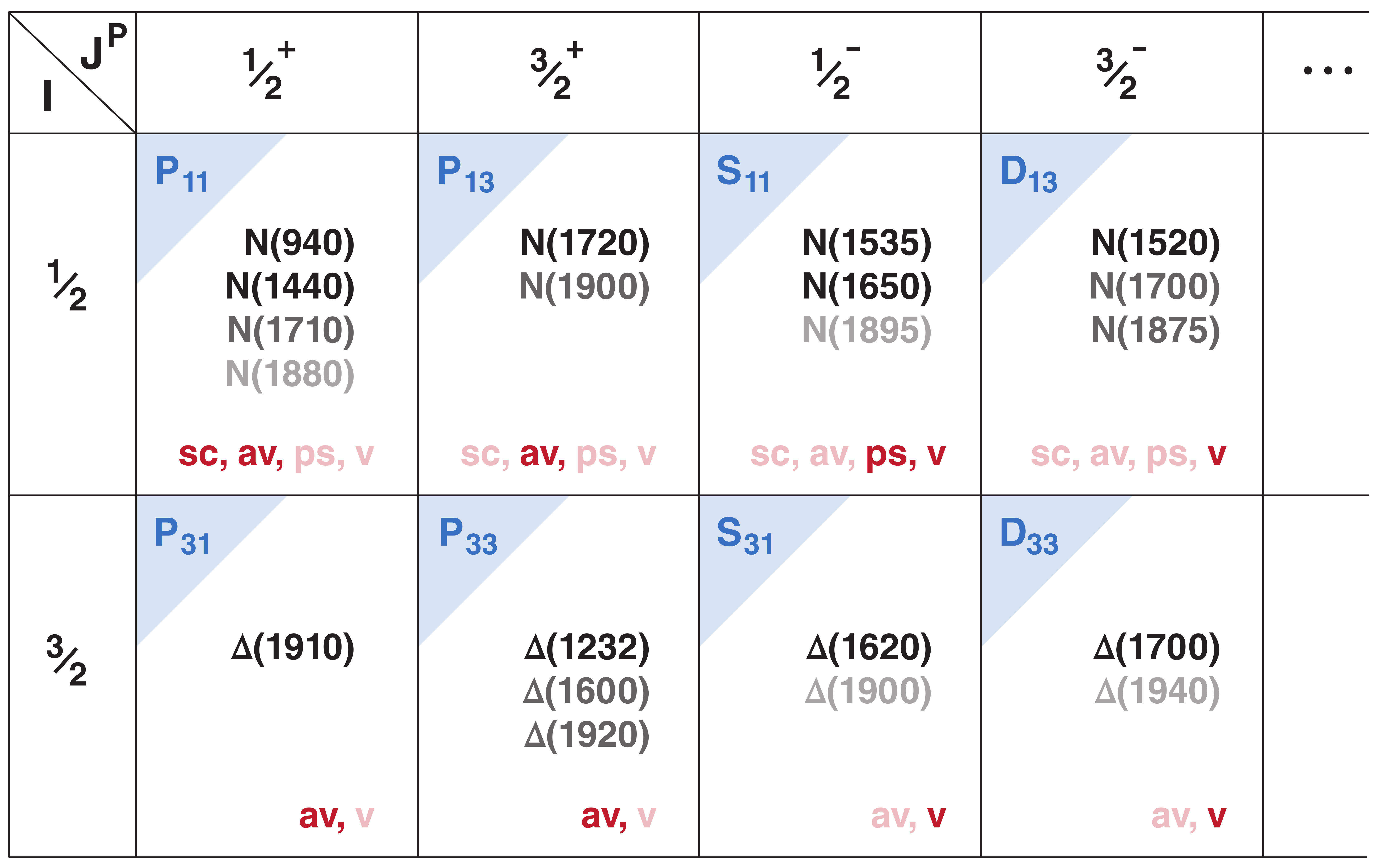}}
            \caption{Nucleon resonances below $2$ GeV~\cite{Agashe:2014kda} and expectations for their dominant diquark content.}
            \label{fig:resonances-table}
            \end{figure*}

 \newpage

 \paragraph{Baryons from quarks and diquarks.} To assess the diquark content for the various nucleon resonances,
          we note that nucleons ($I=\nicefrac{1}{2}$) can feature both diquark isospins whereas $\Delta$ baryons ($I=\nicefrac{3}{2}$) can only consist of $I=1$ diquarks.
          Moreover, baryons with $J=\nicefrac{3}{2}$ should be dominated by $J=1$ diquarks because those with $J=0$ require orbital angular momentum;
          and positive/negative-parity baryons should be dominated by positive/negative-parity diquarks.
            Hence we arrive at Fig~\ref{fig:resonances-table}, which shows the $J^P = \nicefrac{1}{2}^\pm$, $\nicefrac{3}{2}^\pm$ nucleon resonances with at least two stars in the PDG.
            Each box corresponds to a given $I(J^P)$ channel, with one ground state and further radial excitations.
            The presumably dominant diquark channels (scalar, axialvector, pseudoscalar or vector) are shown in bold font.

            To test whether these expectations hold we applied the recipe in Fig.~\ref{fig:quark-diquark} (the details of the calculation will be given elsewhere).
             After solving the quark DSE, the various diquark amplitudes and propagators are calculated
            from their BSEs and quark loop integrals~\cite{Eichmann:2009zx}. They subsequently enter in the quark-diquark BSE from where the baryon masses are determined.
            The results for the ground states are collected in Table~\ref{tab:results}. The \{sc, av, ps, v\} entries provide a measure for the importance of the various diquark channels: they
            denote the magnitude of the quark-diquark amplitudes' dressing functions corresponding to the leading tensor structures at vanishing relative momentum, all normalized to the strongest component.
            So far we have neglected the $0(1^-)$ vector diquark which would only contribute to nucleons but not to $\Delta$ baryons; `v' therefore refers to the $1(1^-)$ diquark only.

            Table~\ref{tab:results} shows that the pseudoscalar and vector diquark contributions are indeed strongly suppressed for the nucleon and $\Delta(1232)$.
            The $N(1535)$, on the other hand, is dominated by pseudoscalar diquarks but
            the other channels (sc, av, v) are all sizeable and similarly important.
            Note that the masses of all states that are dominated by pseudoscalar or vector diquarks are severely underestimated compared to experiment. This is a consequence
            of what was discussed above: in analogy to scalar and axialvector mesons, the mass scales for pseudoscalar and vector diquarks should obtain repulsive shifts beyond rainbow-ladder
            and so would the masses of the baryons that they constitute.
            In the contact-interaction model of Ref.~\cite{Roberts:2011cf} a fictitious coupling strength was introduced into the BSEs for those channels
            to mimic beyond-rainbow-ladder effects and reproduce the $\rho-a_1$ splitting. If we adopted the same strategy here, then Table~\ref{tab:results} suggests that
            such a single parameter could indeed move the $N(1535)$, $\Delta(1620)$, $\Delta(1700)$ and $\Delta(1910)$ masses into the ballpark of their experimental values.
            So far we did not obtain convergent solutions for the $N(1520)$ and $N(1720)$, which could be an artifact of the omission of $0(1^-)$ diquarks,
            or it might also signal a breakdown of the quark-diquark description for these states.

 \vspace{-2mm}

\begin{figure*}[t!]
            \centerline{%
            \includegraphics[width=0.87\textwidth]{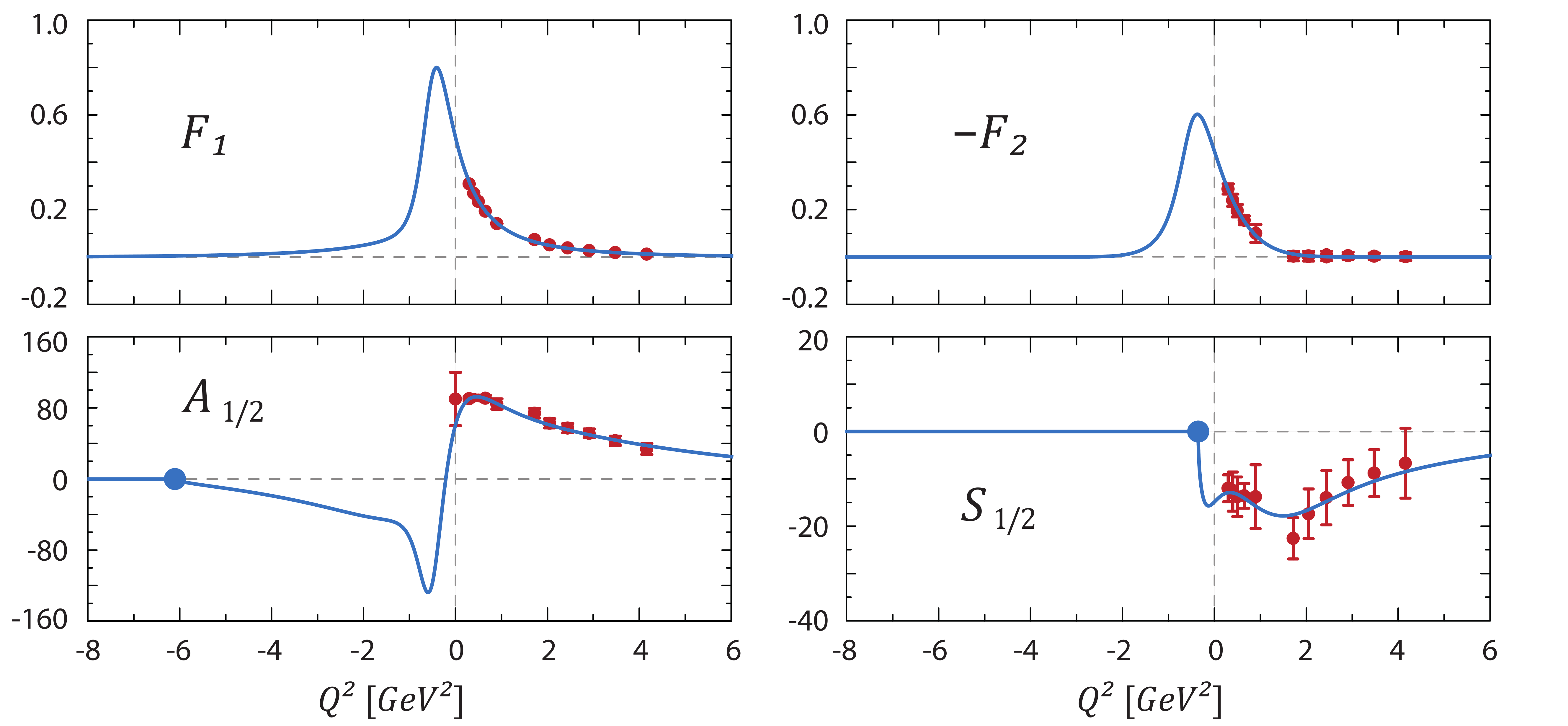}}
            \caption{CLAS data for the $p\gamma^\ast\to N(1535)$ transition form factors and helicity amplitudes~\cite{Aznauryan:2011qj}, together with a toy parametrization 
                     including a $\rho-$meson bump. The helicity amplitudes carry units of $10^{-3}$ GeV$^{-1/2}$. }
            \label{fig:ff-vs-helicity}
            \end{figure*}

 \paragraph{$ N\gamma\to N(1535)$ transition.}
            We conclude by returning to Fig.~\ref{fig:timelike-ffs-1}. The timelike vector-meson structure is contained in the same quark-photon vertex that appears in all current matrix elements,
            so we should expect similar features for nucleon transition form factors. These are traditionally discussed in terms of helicity amplitudes.
            In Fig.~\ref{fig:ff-vs-helicity} we plot the JLab/CLAS data for the $N\gamma^\ast\to N(1535)$ transition form factors and corresponding helicity amplitudes.
            We define the form factors $F_1(Q^2)$ and $F_2(Q^2)$ from
             \begin{equation}\label{current-onshell-resonance}
                 J^\mu =  i\,\conjg{u}(p_f)\left[ \frac{F_1(Q^2)}{m_N^2}\,t^{\mu\nu}_{QQ}\gamma^\nu + \frac{F_2(Q^2)}{2m_N}\,\frac{i}{2} [\gamma^\mu,\slashed{Q}]\right]u(p_i)\,, \qquad
                 t^{\mu\nu}_{AB} = A\cdot B\,\delta^{\mu\nu} - B^\mu A^\nu
             \end{equation}
            to expose the transversality and analyticity properties of the offshell spin-$\nicefrac{1}{2}$ transition vertex~\cite{inprep}  from where the transition matrix element is derived
            (the standard Dirac-like form factor $F_1(Q^2) \,Q^2/m_N^2$ has a kinematic zero at the origin).
            For illustration we parametrize the form factors $F_1$ and $F_2$ by a simple ansatz including a single $\rho-$meson bump.
            Observe that the rich structure of the helicity amplitudes is essentially
            due to kinematic effects, including kinematic zeros at threshold and pseudothreshold $Q^2 = -(m_{N^\ast} \pm m_N)^2$.
            Vice versa, it was noted in Ref.~\cite{Ramalho:2011fa} that the Pauli-like form factor $F_2$ practically vanishes over a wide $Q^2$ domain and only rises at very low $Q^2$.
            This is in contrast to the usual multipole behavior of form factors and rather resembles what one would expect from chiral meson-cloud effects.
            Hence, to access the underlying properties of QCD from the data it is preferable to discuss the behavior of form factors, which are free of kinematic constraints, rather than that of helicity amplitudes.

             \begin{table*}[t]

             \begin{equation*}
             \begin{array}{@{\quad} l @{\quad} || @{\quad\quad} c @{\quad\quad} | @{\quad} c  @{\quad} || @{\quad} c  @{\quad} | @{\quad} c  @{\quad} | @{\quad} c  @{\quad} | @{\quad} c  @{\quad}     }

                                         & N     & N(1535)  &  \Delta(1232)   &  \Delta(1620) & \Delta(1700)  & \Delta(1910)       \\[1mm] \hline\hline \rule{-0.0mm}{0.5cm}

                       \text{Mass [GeV]} & 0.95  &  1.21    &  1.28     &  1.42         &  1.46         & 1.60                \\[1mm]\hline \rule{-0.0mm}{0.5cm}
                       \text{sc}         & 1     &  0.45    &           &               &               &                     \\
                       \text{av}         & 0.37  &  0.56    & 1         & -0.40         & -0.10         & 0.39               \\
                       \text{ps}         & 0.02  &  1       &           &               &               &                     \\
                       \text{v}          & 0.03  &  0.27    & -0.02     &  1            &  1            & 1

             \end{array}
             \vspace{1mm}
             \end{equation*}

               \caption{Rainbow-ladder masses and diquark contributions (see text) for various nucleon resonances.}
               \label{tab:results}

             \end{table*}

 \vspace{-3mm}

  \begin{acknowledgements}
  I would like to thank C. S. Fischer for a critical reading of the manuscript.
  \end{acknowledgements}

 \bibliographystyle{spbasic}


 \vspace{-3mm}


 \end{document}